
\magnification=\magstep1
\hsize=5.9truein
\vsize=8.375truein
\hoffset=.45in
\parindent=25pt
\nopagenumbers
\raggedbottom
\hangafter=1
\def\makeheadline{\vbox to 0pt{\vskip-40pt
   \line{\vbox to 8.5pt{}\the\headline}\vss}\nointerlineskip}
\def\approxlt{\kern 0.35em\raise 0.45ex\hbox{$<$}\kern-0.66em\lower0.5ex
   \hbox{$\scriptstyle\sim$}\kern0.35em}
\def\approxgt{\kern 0.35em\raise 0.45ex\hbox{$>$}\kern-0.75em\lower0.5ex
   \hbox{$\scriptstyle\sim$}\kern0.35em}
\vglue1.5truein
\centerline{\bf A Study of Fermions Coupled to Gauge and Gravitational Fields
on a Cylinder}
\vskip22pt
\centerline{Ralph P. Lano and V.G.J. Rodgers}
\centerline{ Department of Physics and Astronomy}
\centerline{ The University of Iowa}
\centerline{ Iowa City, Iowa~~52242--1479}
\centerline{ Jan.  1994 }
\vglue2.0truein
\baselineskip=12pt
\centerline{\bf ABSTRACT}
\vskip22pt
Fermions on a cylinder coupled to gravity and gauge fields are examined by
studying the geometric action associated with the symmetries of  such a system.
The gauge coupling
constant is shown to be constrained and the effect of gravity on the  masses is
discussed.
Furthermore, we introduce a new  gravitational theory which couples to the
fermions by promoting the
coadjoint vector of the  diffeomorphism sector to a dynamical variable. This
system, together with
the  gauge sector is shown to be equivalent to a one dimensional system.
\vfill
\eject
\headline={\tenrm\hfil\folio}
\baselineskip=16pt
\pageno=1
The study of gauge theories both abelian (Ref.[1]) and non-abelian
(Ref.[2,3]) on a cylinder has been an issue in the literature for some time.
The cylinder provides a toy manifold where one hopes to find insight into
the more pressing 3+1 dimensional theories.
In this note we will take a somewhat different view of fermions
on a cylinder (or a plane) by examining the geometric actions that
arise from the effective action of fermionic theories.  Furthermore we
promote the background gauge fields to propagating fields and
discuss the analogue of this for gravity.

We begin by considering the action
$$
S =   \int d^2 x {\bar \Psi_i}  \gamma^\mu (\partial_\mu \delta_{ij} + q
A^{ij}_\mu ) \Psi_j  + \int d^2 x {\bar \Psi_i} \gamma_\mu \gamma_\nu \Psi_i
T^{\mu \nu} +  {1 \over \alpha} \int F^{\mu \nu} F_{\mu \nu} d^2x,
\eqno(11)$$
where $\Psi$ is a two component Dirac fermion and $A_\mu$ is a gauge field
valued in the vector subalgebra of $U(N) \times U(N)$.  The field $T^{\mu \nu}$
is a symmetric tensor that will serve as a background stress energy source.
We  define a mass term or Yukawa type coupling to the fermions through the
trace of this tensor with the metric.
Since we are considering a coupling to two dimensional gravity we introduce
the zweibein $e^{(a, \alpha)}$ and write
$$
{\bar \Psi_i}  \gamma^\mu \partial_\mu \delta_{ij}  \Psi_j =
{\bar \Psi_i}  \gamma_a e^{(a, \mu)} \partial_\mu \delta_{ij}  \Psi_j.
$$
Lets us write
$\Psi = \pmatrix{ \Psi_A \cr \Psi_B \cr} $ and our gamma
matrices as
$$ \gamma_0 = \pmatrix{0&1\cr 1&0\cr}~~ \gamma_5 = \pmatrix{1&0\cr 0&-1\cr}
{}~~\gamma_1 = \pmatrix{ 0&-1\cr 1&0\cr}
$$
as well as the light cone matrices
$$\gamma_+ = \gamma^-  = \pmatrix{0&2\cr 0&0\cr}~~~ \gamma_- = \gamma^+ =
\pmatrix{0&0\cr 2&0\cr}. $$

With this the fermionic part of the lagrangian may be rewritten as
$$\eqalign{
L =  2 &\Psi_B^* e^{(+,0)}(\partial_0 + q A_0) \Psi_B + i 2 \Psi_B^* e^{(+,1)}
(\partial_1 + q A_1)\Psi_B\cr
 +  2 &\Psi_A^* e^{(-,0)} (\partial_0 + q A_0) \Psi_A
+  2 \Psi_A e^{(-,1)}(\partial_1 + q A_1) \Psi_A\cr
+ 2 & \Psi_A^* e^{(-, \mu)} e^{(+, \nu)} \Psi_B T_{\mu\nu}, } \eqno(2)
$$
where in the above we are using light-cone coordinates for the local
indices and $0$ and $1$ for the coordinate indices.  (This is just our labeling
convention to distinguish the indices.)   Also we have
suppressed the  group indices.  By making the usual change in variables, viz
$\chi_A = {\sqrt 2 e^{(-,1)}} \Psi_A$ and $\chi_B = {\sqrt 2 e^{(+,0)}}
\Psi_B$,
 and fixing the gauge where
$A_0 = 0$ and
$$g^{\mu \nu} = \pmatrix{{e^{(-,0)}\over e^{(-,1)}}&1\cr 1&0\cr} \eqno(3) $$
we finally may write
$$
\eqalign{
S = &\int d^2x ( \chi_B^* \partial_0 \chi_B +  \chi^*_A (\partial_1 + q
A_1)\chi_A
+ \chi_A^* g^{0 0}\partial_0 \chi_A \cr
& \chi_A^* g^{00} T_{00}\chi_B +  \chi_B^* g^{00} T_{00}\chi_A +
 \chi_A^*  T_{01}\chi_B +  \chi_B^*  T_{10} \chi_A).}   \eqno(4)
$$
Since we are in the axial gauge the ghost decouple from the gauge field
sector, while the diffeomorphism ghosts will shift the value of the
central extension in the effective action by 26, Ref.[4,5].
Through to rest of this study we will set $T_{01}=T_{10}=0$.
These terms contribute to
the usual mass term in Minkowski space or Euclidean space.

This action is known to be equivalent to WZNW models and
Polyakov's 2D gravity, Ref.[4,6].  These actions arise as the geometric
 actions
of the  Kac-Moody and Virasoro groups as first discussed in
Refs.[7,8,9].  Since our fermions are coupled to
both the background gauge and the background energy momentum tensor
$T^{\mu \nu}$, we must study
the semi-direct product of the Virasoro and affine Kac-Moody algebras
to extract the effective action.  This is done in Ref.[10].

Let us recapitulate the salient features of that geometric action.
One starts with the algebra
$$
\eqalign{
 [ J^{\alpha}_N, J^{\beta}_M] & = i f^{\alpha\beta\gamma} J^\gamma_{N+M} +
N k \delta_{M+N,0} \delta^{\alpha\beta} \cr
[ L_N,J^{\alpha}_M] & = - M J^{\alpha}_{M+N} \cr
[ L_N,L_M ] &= (N-M) L_{N+M} + {c\over 12} (N^3-N) \delta_{N+M,0}}
$$
where $c = {2 k {\rm Dim}(G)\over 2k + c_v}$, Dim$(G)$ is the dimension
of the group and $c_v$ is the value of the quadratic Casimir
in the adjoint representation.  Since we are interested in dynamics
on a cylinder we explicitly write
$$ \eqalign{ L_N& = i \exp{(i N\theta)} \partial_\theta \cr
J^\alpha_N& = \tau^\alpha \exp{(i N \theta)}.}
$$
and where we have  normalized the generators so that
$Tr( \tau^\alpha \tau^\beta)=
\delta^{\alpha\beta}$.
A basis element for the above algebra would be
$F_{AB}^{\beta}(\rho) = (L_A,J^\beta_B,\rho)$ and  a
generic adjoint element will be written as $F = (\xi (\theta), \Lambda(\theta),
a)$.  The dual space vectors
or coadjoint vectors will be denoted as $B(\theta) = (T(\theta),
A_\theta(\theta),\mu = {1\over q})$ where $q$ is the coupling constant
introduced in Eq.[1].
$\theta $ denotes the circle parameter.
The action of $F$ on $B$ defines the coadjoint representation and we find
that
$$\eqalign{
&\delta_F B \equiv (\xi(\theta),\Lambda(\theta),a) \ast
(T(\theta),A_\theta(\theta),
\mu) = \cr
-&\bigl( 2 \xi'T+T'\xi + {c \mu\over 24\pi} \xi''' - Tr[ A_\theta \Lambda'],
 A_\theta'\xi+A_\theta\xi' + [\Lambda A_\theta - A_\theta \Lambda] +  k \mu
\Lambda', 0 \bigr)\cr
= & (\delta T, \delta A_\theta, 0)}\eqno(5)
$$
where $'$ denotes $\partial_\theta$.
As is well known the space of coadjoint vectors may be foliated in terms of
equivalence classes.  If, say, $B_1$ can be transported to  $B_2$ via a
group transformation then $B_1$ and $B_2$ belong to the same equivalence
class.  These classes are the coadjoint orbits defined by the action of the
Kac-Moody and Virasoro groups on the coadjoint
vectors.  On each orbit there exists a symplectic two form defined by
$$
\Omega_B(B_1,B_2) = \langle B \mid [ F_1,F_2] \rangle,
$$
where here $\langle B \mid F \rangle$ denotes a suitable pairing of coadjoint
vectors with
the adjoint vectors.  The vector $F_1$ yields $B_1$ through $\delta_{F_1} B
=B_1$.
Once this is known then the construction of the geometric action is
straightforward.
 The geometric action associated with fermions coupled to background gauge and
gravitational fields (here we mean the background tensor $T$) is
$$\eqalign{
&S = \int T(\theta) \bigg\lbrack {\partial_\lambda s \over \partial_\theta s}
 {\partial \over \partial \theta} \bigg( { \partial_\tau s \over
\partial_\theta s}\bigg)
-  {\partial_\tau s \over \partial_\theta s} {\partial \over \partial \theta}
 \bigg( { \partial_\lambda s \over \partial_\theta s} \bigg) \bigg\rbrack~
d\lambda d\tau d\theta +
\cr
& \int Tr A(\theta)\bigg\lbrace
 {\partial_\lambda s \over \partial_\theta s}{\partial \over \partial \theta}
(g^{-1}\partial_\tau g)-{\partial_\tau s \over \partial_\theta s}{\partial
\over \partial \theta} (g^{-1}\partial_\lambda g)
+[ g^{-1}\partial_\lambda g,
g^{-1}\partial_\tau g] \bigg\rbrace d\lambda d\tau d\theta
\cr
&+ {c \mu  \over 48\pi } \int  \left[
{{\partial^2_{\theta} s}\over{(\partial_{\theta}s)^2}} \partial
_{\tau} \partial_{\theta} s -  {{(\partial^2_{\theta}s)^2
(\partial_{\tau} s)}\over{(\partial_{\theta} s)^3}} \right] d\theta d\tau
     - {k \mu} \int Tr g^{-1} {{\partial g}\over{\partial
\theta}}g^{-1}
{{\partial g}\over{\partial \tau}}d \theta d \tau
\cr
&+{k \mu} \int Tr g^{-1} {{\partial g}\over{\partial \theta}}
\left[ g^{-1}
{{\partial g}\over{\partial \lambda}}, g^{-1} {{\partial g}\over{\partial
\tau}} \right]
d\theta d\tau d\lambda \cr
 } \eqno(6)
$$
In the above $T(\theta)$ has been induced by the $T^{0 0}$ in the
original lagrangian.
The coordinates $ 0,1 $ are now written as $\tau, \theta$ respectively where
 the  cylindrical
radius, $r$, has been suppressed and is  multiplying $\theta$.
The induced metric component is now written as
$ g_{00} = { \partial_\tau s \over \partial_\theta s} $.   The parameter
$k$ denotes the level of the Kac-Moody algebra.

So far we have only used the fact that  $A_\theta$ and $T$ transform in certain
ways under the time independent gauge transformations and circle
diffeomorphisms.
Recall that $A_\theta$ transforms under gauge transformations as
$$
[ A_\theta, \Lambda ] -  k \mu \Lambda' = \delta A_\theta.
$$
In order that the WZNW action admits a well defined quantum theory, the
coefficient of the Wess-Zumino term must be integer moded.  This would
imply that
$$
k \mu = N
$$ for some integer $N$.  The fact that $k$ has to be an integer arises
independently from the requirement that the representation of the Kac-Moody
algebras be irreducible and unitary.  In fact for the case of the
fermions, we need $k=1$ in order to establish the bosonization
prescription between $ \Psi^i_A \Psi^j_B$ and $ g^i_j(\theta,\tau)$, Ref[6].
 With this we can conclude that the coupling constant $q$ is quantized to
values $1\over N$ for $N$ an integer.   Geometrically this is precisely
the statement that the only orbits that can be related to physical systems are
of the form, (Ref.[11]) $$
B=(T(\theta), A(\theta), N).
   $$
This freezing of the coupling constant suggests that this theory
may be related to a point particle problem.

We now would like to go further by studying the classical equations of motion
of this system and reducing it to a one dimensional system.  The only relevant
degrees of freedom will be the Wilson loops of the fields and
their conjugate momenta.
By integrating by parts and assuming that $T$ and $A_\theta$ have
$\lambda, \tau,$ and $\theta$ dependence  one can show that the first and
second summand in Eq.[6] are equal to
$$
I_T = \int d^2 x~ T(\theta){ \partial_\tau s \over \partial_\theta s} +
\int d^3 x (\partial_\tau T~ { \partial_\lambda s \over \partial_\theta s} -
\partial_\lambda T~ { \partial_\tau s \over \partial_\theta s} ) \eqno(7)
$$
and
$$
I_{A_\theta} = -\int d^2x A_\theta g^{-1}\partial_\tau g + \int d^3x
(\nabla_\lambda A_\theta) g^{-1}
\partial_\tau g - \int d^3x (\nabla_\tau A) g^{-1} \partial_\lambda g
\eqno(8)$$
where
$ \nabla_\tau = \partial_\tau - {\partial_\tau s \over \partial_\theta s}
\partial_\theta - \partial_\theta ( {\partial_\tau s \over \partial_\theta s})
$ and a similar expression for $\nabla_\lambda$.
Now the equations of motion are easy to find when the background
fields $A_\theta(\theta)$ and $T(\theta)$ are included.
The above actions will be used to isolate candidate dynamic terms
for $A_\theta$ and $T$. The equations
of motion for the fermions are equivalent to the equations of motion for the
$s(\theta,\tau)$ (the Weyl components) and $g(\theta,\tau)$ fields which are,
$$
\partial_\theta T(\theta) {\partial_\tau s \over \partial_\theta s}
+ 2 T(\theta) \partial_\theta {\partial_\tau s \over \partial_\theta s} +
{  c \mu \over 24\pi} \partial^3_\theta {\partial_\tau s \over \partial_\theta
s} - Tr [ A_\theta \partial_\theta (g^{-1} \partial_\tau g)] = 0 \eqno(9)
$$
and
$$
(\partial_\theta {\partial_\tau s \over \partial_\theta s}) A_\theta +
\partial_\theta A_\theta ({\partial_\tau s \over \partial_\theta s})
- [ g^{-1}\partial_\tau g , A_\theta ] + k \mu \partial_\theta
(g^{-1}\partial_\tau g ) = 0. \eqno(10)
$$
In the absence of gravity, Eq.[9] has solutions
$g(\theta,\tau) = L(\theta) R(\tau),$ where $L(\theta)$ is arbitrary and
the generators of $R(\tau)$  commute with $A_\theta$.  In other words
$R(\tau)$ belongs to the little group of $A_\theta$.  Since we have used the
time dependent gauge transformations to fix the $A_\theta = 0$ gauge, we must
set $R={\b 1}$.
However, for the sake of demonstration, we leave $R$ unfixed for the moment.
In the presence of gravity this solution is modified to read
$$
g(\theta, \tau) = L(\theta) M(\theta, \tau) R(\tau), \eqno(11)
$$
where $L(\theta)$ is an arbitrary group element,
$R(\tau)$ is again an element of the little group of $A_\theta$, and
$$M(\theta, \tau) = {\rm T} \exp ({-1 \over k\mu} \int_{-\infty}^\tau
{\partial_t s \over \partial_\theta s} A_\theta~ dt), \eqno(12)
$$
which is a time ordered exponential with the boundary condition that
$$g(\theta,\tau\rightarrow -\infty)=1.$$
We notice that ${\partial_\tau s \over \partial_\theta s} A_\theta$ plays the
role of an effective $A_\tau$ component.   With the above
solution, $g^{-1} \partial_\tau g = R^{-1} \partial_\tau R - {1\over k \mu}
{\partial_\tau s \over \partial_\theta s} A_\theta$.  Using this
we may rewrite Eq.[9] as
$$
\partial_\theta {\hat T} {\partial_\tau s \over \partial_\theta s}
+ 2 {\hat T} \partial_\theta {\partial_\tau s \over \partial_\theta s}
 + { c \mu \over 24\pi} \partial_\theta^3 ( {\partial_\tau s \over
\partial_\theta s}) = 0, \eqno(9')
$$
where ${\hat T} = T + {1 \over 2 k \mu} {\rm Tr}[ A_\theta A_\theta]$.
The ${\rm Tr}[ A_\theta A_\theta]$ is a remnant of the Sugawara construction
of the Virasoro generators from the Kac-Moody algebra.
This shows that on shell, the gauge potential in the
presence of gravity modifies the coadjoint vector and may be
thought of as influencing the mass term of the fermions.
More interesting is the fact that ${\hat T}$ is invariant under gauge
transformations whereas $T$ is not (see Eq.(5)).  In general one cannot
expect the addition of rank two tensors to provide one with a new
coadjoint vector.  This is due to the fact that the configuration space for
the coadjoint vectors is not a vector space (although it is a convex space for
fixed central extensions).
However, ${\rm Tr}[ A_\theta A_\theta]$ transforms homogeneously under circle
diffeomorphisms (no central extension), so it can be added to $T$ and the
sum will remain in the configuration space, while the new coadjoint vector,
${\hat T}$,
will assume the same extension.
  Later on we will see that it is ${\hat T}$ that becomes the relevant field in
the
interacting case.

When $T=0$, the symmetries of the semi-classical vacuum
are fully dictated by $A_\theta$ and the central extension.    The solution
to Eq.[$9'$] is simple in that ${\partial_\tau s \over \partial_\theta s}$
must be a linear combinations of the generator in the Virasoro group that
stabilizes ${\hat T}$.
In general the viable orbits are of only a few varieties.   These are
classified for
physics consumption in Ref[12].  These orbits are ${\rm Diff} S^1/S^1$, $
{\rm Diff} S^1/ SL(2,R)_n$, $ {\rm Diff S^1}/T_{(n,\Delta)}$, and
${\rm Diff S^1}/{\tilde T}_{n\pm}$.   For the first two varieties of orbits (we
will call these orbits of the ``first type'')
one can show that they can be identified with constant functions
${\hat T}$.   However the last two orbits are
stabilized by vectors $\cos(n \theta)$ and $1-\cos(n \theta)$ respectively.
The
defining coadjoint vectors are not constants, and are in fact diffeomorphic to
the coadjoint
vectors, $$T_1 = {1\over 2}{ 1\over \cos^2 (n \theta)} + {n^2 c\mu\over 24}~~~
{\rm and}~~~  T_2 = {1\over 2}{1\over (1-\cos(n \theta))^2} +{n^2 c \mu\over
24}$$ respectively.  Notice that these fields are singular at a finite number
of points on the circle.
  Since we are interested in
functions that are not singular on the circle,
 we will ignore the  $ {\rm Diff S^1}/T_{(n,\Delta)},$  and
${\rm Diff S^1}/{\tilde T}_{n\pm}$ type orbits (``second type'').  Therefore
${\hat T(\theta)}$
will be diffeomorphic to a constant function in $\theta$, Ref.[12].
(It is worth noticing that with a suitable choice of $A_\theta$ and $\mu$, one
can
send ${\hat T}$ to an orbit of the first type, even when $T$ defines orbits of
the
second type.)   Thus we will use these constant
coadjoint vectors to represent the orbits.
Let $T_0$ denote the constant coadjoint vector that defines the
orbit on which  ${\hat T}$ lives.  Then (up to an overall scale factor),
$$
{\partial_\tau s(\theta', \tau)\over\partial_{\theta'} s(\theta', \tau)} =
h_1(\tau) + h_2(\tau) \sin ( {\sqrt {24 T_0 \over c \mu }} \theta')
+ h_3(\tau) \cos ({\sqrt {24 T_0 \over c \mu }} \theta'), \eqno(13)$$
where $\theta'$ is related to $\theta$ by that diffeomorphism that
carries ${\hat T}$ into $T_0$.  From now on we will drop the distinction
between
$\theta'$ and $\theta$.
The functions $h_1, h_2,$ and $h_3$ are arbitrary time
dependent real functions and together they generate an SL(2,R) symmetry.
Notice, however, that Eq.[9'] is homogeneous in $s(\theta, \tau)$.  This
is a remnant of  Weyl symmetry on the metric.
Thus one can eliminate one of the $h$ functions.  Also we should keep in mind
that time reparameterization invariance was used to fix the gauge.
Furthermore the quantity
$R \equiv \sqrt{h_1^2-h_2^2-h_3^2}$ is an SL(2,R) invariant.  Since we will not
allow any time dependence in this invariant, we may
consider solutions to Eq.[9']  were $h_1, h_2, $ and $h_3$ are constants.
With this, the solution of $s(\theta, \tau)$  is  any function of the
parameter $z$, viz. $s(z)$, where $z$ is given by
$$
z = \tau + {2 c \mu\over\sqrt{24\pi T_0 (h_1^2-h_2^2-h_3^2)}}  {\rm
ArcTan}({(h_1 - h_2) {\rm Tan}( {\sqrt {24\pi T_0 \over c \mu }} \theta) +h_3
\over \sqrt{h_1^2-h_2^2-h_3^2}}). \eqno(14)
$$
There are three distinct regions for the SL(2,R) invariants corresponding to
$R^2>0, R^2 = 0,$ and $R^2 < 0$.  Let us examine these three regions.

 For $R^2 >0$, we may choose, $h_1 = R, h_2=h_3 =0$.  Then
$z$ becomes
$z = \tau + {\theta \over R}$.  This is just the usual light cone coordinate
with $\theta$ rescaled by $R$.  Notice that there is no  $T_0$ dependence in
$s$ for this particular choice of parameters.  This sector will be our
preferred choice. Therefore we will require that $h_1^2 > h_2^2 + h_3^2$.

 In the $R^2 = 0$ sector, there are several different directions that
$z$ may go.  Clearly for $h_1=h_2=h_3=0$, $z=\theta$.  However other
possibilities
exists.  The cases when $h_1 = \pm h_3 = h,$ and $ h_2=0$ or when
$h_1 =\pm h_2 =h $ and $h_3 =0$ , both correspond to ${\partial_\tau s(\theta,
\tau)\over\partial_\theta s(\theta, \tau)}$ stabilizing an orbit of the second
type.
 For this reason we will not be concerned with the
$R^2 = 0$ sector.

And finally for the $R^2 < 0$ sector, again we find that for $h_1 = h_3 =0$ and
$h_2 \neq 0,$ that ${\partial_\tau s(\theta, \tau)\over\partial_\theta
s(\theta, \tau)}$ stabilizes orbits of the second type  when $h_2$ is real.

With all of this in place, one may now add dynamics to {\it both} the
$A_\theta$ and $T$ fields.
This will allow both the background gauge and ``stress-energy tensor''
to propagate.   We will then try to find  one dimensional systems that
correspond to our action.  To proceed, we will consider the ``free''
theories first and then add the fermion interactions.
A suitable choice of kinetic terms for both the $A_\theta$ and $T$ is
required.
The action, Eq.[6], naturally contains the
 gauge fixed Cherns-Simons (when $g^{-1}\partial g$ is included with $A$)
terms for both the gauge and diffeomorphism sectors.  This is seen
by examining Eqs.[7] and [8].   (In Eq.[8] only the ordinary part of the
covariant derivative contributes to the Chern-Simons term.)  Instead
of using such topological actions
 that are restricted to  certain dimensions,
 we would like to provide dynamics to $A_\theta$ and $T$ through
curvature squared type actions.  We can use the
Chern-Simons terms to identify
the curvatures that we need to construct the two dimensional actions.
The dynamics for $A_\theta$ follows from ${1\over \alpha} F^{\mu\nu}
F_{\mu\nu}$ so we will add to the action the lagrangian
$$
{1\over \alpha} \partial_\tau A_\theta \partial_\tau A_\theta
$$
and we will drop the Chern-Simons term,
$$
 \int d^3x (\partial_\lambda A) g^{-1} \partial_\tau g - \int d^3x
(\partial_\tau A) g^{-1} \partial_\lambda g
$$ from consideration.  Similarly, by examination of the Diff  Chern-Simons
term,
we can pick off the appropriate curvature for $T$ in the absence of $A_\theta$.
 This term  is covariant under the
residual Diff transformations and write
$$
I_T = \int d^2 x~ T(\theta){ \partial_\tau s \over \partial_\theta s} + {1\over
\beta} \int d^2 x \partial_\tau T \partial_\tau T, \eqno(15)
$$
where $\beta$ has units of ${\rm mass}^4$.
This action will replace Eq.[7].  This use of kinetic terms allows us to
deviate
from conformal field theories and should also be contrasted with methods used
to construct model spaces for the Kac-Moody and Virasoro algebras such as those
found in  Ref[13,14].

We can now introduce the conjugate variable $E$ to $A_\theta$, keeping in
mind the Gauss' law constraint that identifies the appropriate $E's$ with
$A's$.  Similarly the conjugate variable $E_T$ is introduced for $T$.
Since the Gauss' laws are the generators of the  time independent
transformations, we can use the isotropy equations from the
coadjoint orbits to identify their form.
For the gauge sector we will follow Ref.[2].
It is instructive to first consider the pure
cases for $A_\theta$ and $T$.

In the case of $A(\theta)$, one ``rotates'' $A_\theta$ to
zero via the transformation,
$\partial_\theta S + A_\theta S = 0.$ Then the Wilson loop, $$S(2 \pi)
= P \exp(-{1\over k\mu}\int\limits_{\theta=0}^{\theta=2\pi}A_\theta d\theta)
S(0) \eqno(16)$$
is identified with the conjugate variable $q$ and one extracts the
associated conjugate momentum through
the Gauss' law constraint $\partial_\theta E + {1\over k\mu} [A_\theta,E] = 0$,
Ref.[2].
Through this one finds that $E(\theta) = S(\theta) E(0) S^{-1}(\theta)$
so that $E(0)$ is the appropriate conjugate momentum, $p$.
The Hamilton equations for $q$ and $p$ follow from the equations of motion
and
$ \partial_\tau ( \partial_\theta S + A_\theta S)=0$.
One has
$$ q^{-1} {\dot q}  = -{2 \pi \over k \mu} p ~~~{\rm and} ~~~~{\dot p} = 0 .$$
The pure Diff (or pure gravity) sector may be treated in an analogous way.
$Q$ and $P$
denote the conjugate variables in this sector.
First the action can be written as
$$ S_T = {1\over \beta} \int \partial_\tau T E_T d^2 x -{1\over 2\beta} \int
E_T E_T d^2 x . $$  The
equation of motion and the Gauss' law constraint for this system
are given by $$\partial_\tau E_T = 0 ~~~{\rm and} ~~~ 2 \partial_\theta E_T T +
\partial_\theta T E_T + t \partial^3_\theta E_T = 0. \eqno(17)$$
 Let us define the function $v(\theta)$ through
$$T(\theta) = t\{\theta,v\}=  t ({\partial^3_\theta v \over \partial_\theta v}
- {3\over 2} ({\partial^2_\theta v \over \partial_\theta v})^2), \eqno(18)$$
where $t = {c \mu \over 24\pi}$ and $\{ , \}$ is the Schwartzian derivative.
Then the Wilson loop for the Diff sector is
$Q = v(2\pi)$.
Using  Eq.[17] we  find the conjugate momentum in terms of $v$ as  $
E(\theta)_T = {E_{\rm T}(0) \over \partial_\theta v},$
 where  $E_{\rm T}(0) = P$.
Since   $T$ can be taken as a constant, we have
$$v(\theta) = \exp( \pm i {\sqrt {2 T\over t}} \theta). \eqno(19)$$
  Now the Hamilton equations for $Q$ and $P$ follow from the equations
of motion and by taking a time derivative of Eq.[18] one can show that
$$ E_{\rm T}(0) ({1\over {\partial v \over \partial \theta}})^3 =
\partial^3_v (\partial_\tau v), \eqno(20)$$
 where $\partial_v = {1\over {\partial v \over \partial
\theta}} {\partial \over \partial \theta}$.
Then $\partial_\tau v$ has the solution,
$$ \partial_\tau v(\theta) = E_T(0) (\int\limits_0^\theta
{v^2(\phi)\over 2(\partial_\phi v(\phi))^2} d\phi
- v(\theta) \int\limits_0^\theta {v(\phi) \over (\partial_\phi v(\phi))^2}
d\phi
+ v^2(\theta) \int\limits_0^\theta {1\over 2(\partial_\phi v(\phi))^2} d\phi )
{}.
$$

 From here on we will consider only the case were $T$ is
a constant.   As remarked earlier this excludes specific orbits.  The solution
to  Eq.[20]
 ( evaluating $v$ at $2 \pi$) is   $$ {\dot Q} =  {2 \pi^3 P \over \log(Q)^3}
(3 - 4Q
+ Q^2 +2 \log(Q)),\eqno(21 a)$$ while the other Hamilton equation comes from
the equations of motion
and is $$ {\dot P} = {P (\log(Q) +1) {\dot Q} \over \log(Q) Q}. \eqno(21 b) $$
The components of the symplectic two form $\omega$ may be written as
$$ \omega_{Q,P} = {  \log^3(Q) \over 2 \pi^3 P^2 (2 \log(Q) + Q^2 -4 Q^2
+3)}$$and the Hamiltonian for this system is
$$
H_o = \log(P) + \log(Q) + \log(\log(Q)). \eqno(22)
$$
Note that $\omega$ is invertible in the region $0 < Q < \infty$.
In particular $\omega \rightarrow {3\over 2}$ as $Q \rightarrow 1$.  Thus
Poisson brackets
are well defined on the phase space.  However the constant energy surfaces
on the phase space reveals forbidden regions for certain initial
data.  Figure 1, shows
characteristic constant ``energy'' surfaces.  There is a clear
bifurcation in the $+$ versus $-$ solutions for $Q$ in Eq.[19].  The positive
set of maps is bounded from below by the $P = 0$ surface while the
negative set is bounded from above by $P= - \exp(1 + E_o)$ where $E_o$ is the
energy.  This barrier is actually welcomed since there was a twofold covering
of the configuration space due to the two types of solutions.
For the case when ${T \over t}$ is positive definite the equations of motion
remain the same but $Q$ is valued on the complex circle.  Specifically,
$$Q = \exp(\pm 2\pi i \sqrt{ | {2 T \over t} |} ), ~~
P = \exp(E_o) \exp\{\mp i (2\pi{2 T \over t} + {\pi\over 2})\} ~2\pi(\sqrt{|{2
T \over t}|} ).$$  Thus $P$ lags behind  $Q$ on the
complex plane by $90^\circ$.  Now the $+$ and $-$ solutions correspond to
clockwise
and anti-clockwise rotation about the complex plane as $ 2\pi(\sqrt{| {T \over
t}|})$ goes from $0$ to $\infty$.

As a side note, for a particular choice of quantum systems, the eigenstates of
the Hamiltonian satisfy the
equation
$$
(i {\partial \over \partial Q} + {\exp(E_o) \over Q \log(Q)})\psi(Q) = 0.
\eqno(23)
$$ The solutions for this case are easily to find and one has that
$$
\psi_{E_o}(Q) = A \exp(i \exp(E_o) \log(\log(Q))).
$$ Of course this quantization is not unique since there will be ambiguities
when one promotes the Hamilton equations to operators as well as the way
we impose the gauge fixing conditions.  For a detailed study of quantization
ambiguities in the Yang-Mills case see Ref.[15].

Let us continue with our preliminary study of this gravitational (Diff)
theory by introducing the fermions sources for the metric.
Here we consider the action given by Eq.(15) together with the
equations of motion for ${\partial_\tau s \over \partial_\theta s}$ when $
A_\theta = 0$.  The equations of motion and the Gauss' constraint
are
$${1\over\beta} \partial_\tau E_T = h_1 + h_2 \sin (\alpha \theta)
+ h_3 \cos (\alpha \theta) ~~~{\rm and} ~~~ 2 \partial_\theta E_T T +
\partial_\theta T E_T + t \partial^3_\theta E_T = 0, \eqno(24)$$
where $\alpha = \sqrt{2 T \over t}$.  By evaluating the above expressions at
$\theta = 2\pi$, the Hamilton equations become
$$\eqalign{
&{\dot P} = P {{\dot Q} \log(Q) + {\dot Q} \over Q \log(Q)} +
{\beta \over 2 \pi} Q \log(Q) (h_1 + {h_3\over 2} (Q+Q^{-1}))\cr
&{\dot Q} = {2 \pi^3 P \over \log^3(Q)}(3 - 4 Q + Q^2 +2 \log(Q)).}\eqno(25)$$
where the reality condition has forced $h_2 = 0$.
The Hamiltonian may be written as
$H=H_o + H_i $, where $H_o$ is given by Eq.(22) and
the interacting Hamiltonian is
$$H_i = {\beta \over 4 \pi^4 P^2} \int\limits_{0}^{Q} { Q'\log^4(Q')
(h_1 + {h_3\over 2} (Q'+Q'^{-1}) \over (2 \log(Q') +3 -4 Q' +{Q'}^2)} dQ '.
$$

Now we consider the fully interacting gauge and gravitational case
with the fermionic sources, $s(\theta, \tau)$ and $g(\theta, \tau)$.
The second term of the action in Eq.(15) is inadequate
since it is not invariant under the transformation, $ \delta_\Lambda T =
 {\rm Tr}[A_\theta \Lambda']$.  However this is remedied by
writing the kinetic term for $T$ as
$$
{\rm I}_T =  \int d^2 x~ T(\theta){ \partial_\tau s \over \partial_\theta s} +
{1\over \beta} \int d^2 x \partial_\tau (T +{1\over 2 k \mu}{\rm Tr}[A_\theta
A_\theta]) \partial_\tau (T +{1\over 2 k \mu}{\rm Tr}[A_\theta A_\theta]).
\eqno(26)$$
For completeness, we write the full action that is being study as,

 $$\eqalign{
&S = \int d^2 x~ T(\theta){ \partial_\tau s \over \partial_\theta s} +
 {1\over \beta} \int d^2 x \partial_\tau (T +{1\over 2 k \mu}{\rm Tr}[A_\theta
A_\theta])
E_T + \cr
&-\int d^2x A g^{-1}\partial_\tau g - \int d^3x \partial_\theta({
 \partial_\lambda s \over \partial_\theta s} A) g^{-1} \partial_\tau g + \int
d^3x \partial_\theta({ \partial_\tau s \over \partial_\theta s} A) g^{-1}
\partial_\lambda g
\cr
&+ {c \mu  \over 48\pi } \int  \left[
{{\partial^2_{\theta} s}\over{(\partial_{\theta}s)^2}} \partial
_{\tau} \partial_{\theta} s -  {{(\partial^2_{\theta}s)^2
(\partial_{\tau} s)}\over{(\partial_{\theta} s)^3}} \right] d\theta d\tau
     - {k \mu} \int {\rm Tr} g^{-1} {{\partial g}\over{\partial
\theta}}g^{-1}
{{\partial g}\over{\partial \tau}}d \theta d \tau
\cr
&+{k \mu} \int {\rm Tr} g^{-1} {{\partial g}\over{\partial \theta}}
\left[ g^{-1}
{{\partial g}\over{\partial \lambda}}, g^{-1} {{\partial g}\over{\partial
\tau}} \right]
d\theta d\tau d\lambda \cr
 &+  {1\over \alpha}  \int d^2 x \partial_\tau A_\theta E -{1\over 2\beta }
\int E_T E_T~ d^2~x -{1\over 2 \alpha} \int E E d^2~x }\eqno(27)
$$
The Gauss' constraints together with the equations of motion for this action,
$$
\eqalign{
&{\rm (a)} ~~~ {1\over \beta}(2 \partial_\theta E_T T + \partial_\theta T E + t
\partial^3_\theta E_T)  -{1\over\alpha}{\rm Tr}[A_\theta \partial_\theta E] = 0
\cr
&{\rm (b)}~~~ \partial_\theta E_T A_\theta + \partial_\theta A_\theta E_T +
[A_\theta, E] +
k\mu \partial_\theta E = 0\cr
&{\rm (c)}~~~ {1\over\alpha} \partial_\tau E = {-E E_T \over \beta k \mu} +
g^{-1} \partial_\tau g
\cr &{\rm (d)}~~~ {1\over \beta} \partial_\tau E_T - {\partial_\tau s(\theta,
\tau) \over
\partial_\theta s(\theta, \tau)} =0 \cr
&{\rm as~ well~ as,} ~~~~~ \partial_\tau A = E ~~~{\rm and}~~~
\partial_\tau (T +{1\over 2 k \mu}{\rm Tr}[A_\theta A_\theta]) = E_T, }
\eqno(28) $$
can now be used to reduce the system to a one-dimensional problem.
The Gauss' law, Eq.[28 b], is satisfied by
$E(\theta) = S p S^{-1} - {\alpha \over \beta k \mu} E_T A_\theta$, where $S$
is the
Wilson line as in the pure gauge case.  This solution together
with Eq.[28 c] and Eq.[28 d] imply that
${\dot p} = 0$.  Here we have used the equation of motion for $g$ and have
written
$g^{-1}\partial_\tau g = -{1 \over k \mu}A_\theta {\partial_\tau s \over
\partial_\theta s}$.
The constraint, Eq.[28 a], reduces to,
$ 2 \partial_\theta E_T {\hat T} + \partial_\theta {\hat T} E_T
+ t \partial^3_\theta E_T =0$, where ${\hat T} = T +{1\over 2 k \mu}{\rm
Tr}[A_\theta A_\theta]$
is the gauge invariant coadjoint vector.  Using the same procedure of the pure
Diff sector,
we will bring ${\hat T}$  to a constant. This tacitly assumes that the
gauge potential $A_\theta$ is well defined on the circle and free from
singularities.  Then the one dimensional system can be shown to
be defined by the Hamilton equations,
$$
\eqalign{
{\dot Q}& =  {2 \pi^3 P \over \log^3(Q)} (3 - 4Q+ Q^2 +2 \log(Q))\cr
 {\dot P}& = {P (\log(Q) +1) {\dot Q} \over \log(Q) Q} +
{\beta \over 2 \pi} Q \log(Q) (h_1 + {h_3\over 2}(Q + Q^{-1}))\cr
 {\dot q}& = -{2 \pi \over k\mu} q p \cr
 {\dot p}& = 0 }\eqno(29)
$$
where the Hamiltonian is
$$\eqalign{
H= & {1\over 2 k \mu} p^2 + \log(P) + \log(Q) + \log(\log(Q)) + \cr
+& {\beta \over 4 \pi^4 P^2} \int\limits_{0}^{Q} { Q'\log^4(Q')
(h_1 + {h_3\over 2} (Q'+Q'^{-1}) \over (2 \log(Q') +3 -4 Q' +{Q'}^2)} dQ'.}
$$

The point particle action is
$$
S = \int ({-\log^3(Q) {\dot Q} \over 2 \pi^3 P (2 \log(Q) +3 -4 Q+ Q^2)} +
pq^{-1}{\dot q}
- H )d~t
$$
Here $q$ and $Q$ are the gauge and Diff Wilson loops respectively, while
$p$ and $P$ are their conjugate momenta.   The parameters $h_1$ and $h_3$
may be used to fix the SL(2,R) invariant parameter, $R$, which will label
the states.

In conclusion, we have shown that the geometric action reveals that the
coupling constant  for fermions coupled to gauge potentials on a cylinder
must be quantized.   Further, we have shown that one may define a new,
albeit gauge fixed,
gravitational type action by looking at the diffeomorphism sector of
the fermions.  The new action is such that it is not restricted to a particular
dimension and in two dimensions is equivalent to a point particle problem. The
Wilson loops, which are
gauge invariant quantities, are the  only dynamical degrees of freedom.
Although the symplectic structure is well defined throughout the
full range of the configuration space, there is a  bifurcation in equal energy
surfaces  due to an ambiguity in assigning the Wilson loop
to a particular coadjoint orbit.
The fully interacting case is examinied and  the gravitational sector
introduces a
scale through the coupling constant $\beta$.   The SL(2,R) invariant
$R$ labels the states.   Since the geometric action can be extended to
the line R$^1$, one suspects that this work can be generalized.  In that
case the coupling constants $\beta$ and $\alpha$ would provide two
new scales to the theory, whereas for the cylinder the scale is fixed by
the radius of the cylinder.

Acknowledgements:  We would like to thank Y. Meurice and N.~Louie
for discussion.   This note was completed while VGJR was visiting the
Mathematical Sciences Research Institute.  This work was supported in part
by NSF grants DMS-9022140 (MSRI) and PHY-9103914.

\vfill\eject
\nopagenumbers
\baselineskip=13pt
\centerline{\bf REFERENCES}
\settabs 1 \columns
\+ [1] J.E. Hetrick and Y. Hosotani, Phys. Rev. D38 (1988) 2621; Phys. Lett.
B230 (1989) 88 \cr
\+ [2] S.G. Rajeev, Phys. Lett. B212 (1988) 203 \cr
\+ [3] E. Langmann and G.W. Semenoff, Phys. Lett. B296 (1992) 117 \cr
\+ [4] A.M. Polyakov, Mod. Phys. Lett. A2 (1987) 893 \cr
\+ [5] V.G. Knizhnik, A.M. Polyakov, and A.B. Zamolodchikov, Mod.Phys. Letts A
3 (1988) 819 \cr
\+ [6] E. Witten,  Nucl. Phy. B233 (1983)  \cr
\+ [7] B. Rai and V.G.J. Rodgers, Nucl Phys. B341 (1990) 119 \cr
\+ [8] A. Yu. Alekseev and S.L. Shatashvili,  Mod. Phys. Lett. A3 (1988) 1551
\cr
\+ [9] P.B. Wiegmann, Nucl Phys. B323 (1989) 311 \cr
\+ [10] R.P. Lano and V.G.J. Rodgers, Mod. Phys. Lett. A7 (1992) 1725 \cr
\+ [11] A. Pressley and G. Segal, pg.(45), {\it Loop Groups}, Oxford Univ.
Press, Oxford, 1986 \cr
\+ [12] E. Witten, Comm. Math. Phys. 114 (1988) 1 \cr
\+ [13] Ho-Seong La, Philip Nelson, A.S. Schwarz, Comm.Math.Phys.134 (1990) 539
\cr
\+ [14] A. Yu Alekseev and S.L. Shatashvili, Nucl. Phys. B323 (1989) 719 \cr
\+ [15] L. Chandar and E. Ercolessi, Syracuse University Preprint SU-4240-537
\cr
\+ ~~~ ``Inequivalent quantizations of Yang-MIlls theory of a cylinder.'' \cr
\end